\documentclass[12pt]{article}
\usepackage{cite}
\usepackage{ulem}
\usepackage{color}
\usepackage{epsfig}
\usepackage{graphicx}
\usepackage{subfigure}
\usepackage{ulem}
\usepackage{amsfonts}
\usepackage{epsfig,bm}
\usepackage{graphicx}
\usepackage{comment}
\newcommand{\be}{\begin{equation}}
\newcommand{\ee}{\end{equation}}
\newcommand{\ba}{\begin{eqnarray}}
\newcommand{\ea}{\end{eqnarray}}

\begin{document}
\begin{titlepage}
\title{Geometric Design and Stability of Power Networks}
\author{}
\date{Neeraj Gupta$^{a}$ \thanks{ngtaj.iitk@gmail.com},\ Bhupendra
Nath Tiwari $^{b}$ \thanks{\noindent bntiwari.iitk@gmail.com} and
Stefano Bellucci $^{b}$ \thanks{\noindent bellucci@lnf.infn.it}\\
\vspace{0.5cm} $^a$Department of Electrical Engineering,\\ Indian
Institute of Technology Kanpur- 208016, India.\\ \vspace{0.5cm}
$^{b}$INFN-Laboratori Nazionali di Frascati\\
Via E. Fermi 40, 00044 Frascati, Italy.}

\maketitle \abstract{From the perspective of the network theory,
the present work illustrates how the parametric intrinsic
geometric description exhibits an exact set of pair correction
functions and global correlation volume with and without the
inclusion of the imaginary power flow. The Gaussian fluctuations
about the equilibrium basis accomplish a well-defined,
non-degenerate, curved regular intrinsic Riemannian surfaces for
the purely real and the purely imaginary power flows and their
linear combinations. An explicit computation demonstrates that the
underlying real and imaginary power correlations involve ordinary
summations of the power factors, with and without their joint
effects. Novel aspect of the intrinsic geometry constitutes a
stable design for the power systems.} \vspace{1.5cm}

\vspace{2mm} {\bf Keywords}: Correlation; Geometry; Power Flow;
Network; Stability.
\end{titlepage}
\section{Introduction}

Planning issues towards a stable power supply, viz., transmission
and distribution systems have been in wide application of the
power system since a decade \cite{ss}. In the transmission theory,
this shows that the power flow varies as a function of the power
factor of the network. In order to regulate the voltage of the
network, the power factor $\phi$ is thus related to the impedance
angle $a \ge \frac{\pi}{2}-\phi$. Interestingly, the determination
of $a$ can be accomplished by tuning the network parameters, viz.,
resistance (r), inductance (L) and capacitance (C) \cite{Miller1,
Miller2}. An optimal choice of $a$ improves the efficiency of
power networks, and thus an appropriate design of the network
parameters. For a given transmission line(s), we provide a proper
network planning for the operations with a minimum reactive power
requirement. From the outset of the intrinsic geometry, the
present paper determines the required unit of the power flow under
the fluctuations of the power factors.

Up to now, most of the network planning and design is based on the
power flow equations \cite{ref1,ref11}, and thus the network
characterizations are linearly afforded by a set of chronological
data analysis, heuristics methods, parametric estimations and
optimization techniques \cite{ref2,ref3}. In this paper, this
motivates us to define non-linear criteria for the electrical
networks. Subsequently, we take an account of the fact that the
set of voltages at all buses attains an equilibrium configuration,
therefore the extremization of the power flow in the network can
be determined in terms of $a$. The selection of the network
parameters determined by the optimization techniques could be
unreliable and thus the cause of bottle-necking. This shows an
urgent need for compensator(s). For a safe operation and optimal
power flow, our analysis provides stability criteria for the
non-linearly reliable comportment of the electrical network. In
this concern, our method provides a high degree of compensation
strategy to reduce the fluctuation effects of the network
parameters, viz., $r$, $L$ and $C$.

To be specific, we focus our attention on the power networks and
determine the required set of voltage stability criteria, viz.,
the selection of power factors, network planning and compensation
strategies. Further, the intrinsic geometry offers an effective
determination of the power system characterization, power factor
corrections and voltage regulation. It is worth mentioning that
the voltage level at the buses could be tuned in the equilibrium.
In the reverse engineering, the proposed solution keeps the power
system in a stable voltage range, under the fluctuations of the
network configuration. In this way, we provide an efficient power
system characterization, which is non-linearly stable over the
fluctuations of the network phases.

In this concern, the intrinsic geometry has been important in the
configurations involving black holes in string theory
\cite{9601029v2,9504147v2,0409148v2,9707203v1,
0507014v1,0502157v4,0505122v2} and $M$-theory \cite{0209114,
0401129,0408106,0408122}, possessing a set of rich stability
structures \cite{0606084v1,SST,bnt, BNTBull,
BNTBull08,BNTBullcorr, Hawking, more1,more2}. Therbay, there have
been several investigations about the equilibrium perspective in
black holes, explicating the nature of pair correlations and the
associated stability of the solutions \cite{RuppeinerPRD78}.
Besides several general notions analyzed in condensed matter
physics \cite{RuppeinerRMP,RuppeinerA20,RuppeinerPRL,RuppeinerA27,
RuppeinerA41}, we consider specific electrical network(s). As
mentioned above, we analyze the parametric pair correlation
functions and their correlation relations about the equilibrium
configuration. We find that the intrinsic geometric consideration
entails an intriguing feature of the underlying fluctuations,
which are defined in terms of the network parameters.

Given a definite covariant intrinsic geometric description of the
network configuration, we expose (i) the conditions for the
stability, (ii) properties of the parametric correlation functions
and (iii) scaling relations in terms of the parameters of the
network. In this analysis, we enlist the complete set of
non-trivial parametric correlation functions of the electrical
networks. This anticipation follows from similar considerations in
the black hole solutions in general relativity
\cite{gr-qc/0601119v1,gr-qc/0512035v1, gr-qc/0304015v1,
0510139v3}, attractor black holes \cite{9508072v3,9602111v3,
new1,new2,0805.1310,bfm1,bfm2,bfm3,bfm4,bfm5,bfm6,bfm7,bfm8,bfm9}
and Legendre transformed finite parameter chemical configurations
\cite{Weinhold1, Weinhold2}, quantum field theory and associated
Hot QCD backgrounds \cite{BNTSBVC,talk}, as well in the stability
of Quarkonia states \cite{Quark}. Thus, the intrinsic geometry
plays an important role in the study of power networks and their
stable design applications.

In the context of power flow equations \cite{ref1,ref11}, the
usefulness of the present investigation may apparently seem to be
limited, however, we illustrate that our preposition could in
principle be generically applied to all electrical networks, in
order to achieve the best understanding of the phenomenon and
importance of the controlled power flow and network analysis. The
intrinsic geometric consideration is capable to provide strategic
planning criteria for the effective use of power system and
voltage stability. We show that this notion follows from the
standard laws of the electrical circuits \cite{ref11}.
Subsequently, for an additional component, the criteria of the
voltage stability can be used for an optimal selection of the
network parameters.


\section{Phases of the Electrical Network}

In this section, we set up the formulation of the problem and
outline the notion of the intrinsic Riemannian geometry. The
exploration of the power flow equations gives the relation of the
power flow with network phases, and thus the consideration for the
analysis. Most efficiently, our intrinsic geometric model is
designed to provide the critical values of the phases. It is worth
mentioning that the present method is important towards the
determination of the parameters, and thus the unstable mode of the
network.

\subsection{Hypothesis of the Power Flow}

For the optimization of the power flow, we share the hypothesis of
the prior solutions and use the load flow equations
\cite{ref1,ref11} in order to solve the mentioned issues. The
associated power conservation equations with the real (resistive)
and imaginary (reactive) branch parameters are given by
\begin{scriptsize}
\begin{eqnarray} \label{gerenalrealpower}
P_i&=& \sum |V_i ||V_j ||Y_{ij} |(G_{ij}  cos(a_{ij}+\delta_j-\delta_i)+ B_{ij} sin(a_{ij}+\delta_j-\delta_i)) \nonumber \\
\end{eqnarray}
\end{scriptsize}
\begin{scriptsize}
\begin{eqnarray} \label{generalimaginarypower}
Q_i&=& \sum |V_i ||V_j ||Y_{ij} |(G_{ij}
sin(a_{ij}+\delta_j-\delta_i)- B_{ij}
cos(a_{ij}+\delta_j-\delta_i))
\end{eqnarray}
\end{scriptsize}
In the above equations, the phases are defined as
\begin{scriptsize}
\begin{eqnarray}
tan(a_{ij})=\frac{X_{L_{ij}}- X_{C_{ij}}}{r_{ij}}
\end{eqnarray}
\end{scriptsize}

The inverse set of the impedances $Z_{ij}$ and voltage angles
$\delta_j$ are

\begin{scriptsize}
\begin{eqnarray}
Y_{ij}&=& \frac{1}{( r_{ij}+ j X_{L_{ij}}- j X_{C_{ij}})} \nonumber \\
\delta_j&=& \frac{V_j}{|V_j|}
\end{eqnarray}
\end{scriptsize}

For the purpose of subsequent analysis, let us consider arbitrary
$i^{th}$-bus such that the steady state condition is realized as
$|V_i |=1$. This makes the underlying configuration reach an
equilibrium. Thus, the respective cases of the present interest
reduce to the standard network considerations. For the purpose of
the future analysis, we consider that a loss-less line is defined
by $a_{ij}= ±90^{0}$, where $+90^{0}$ represents the ideal case.
In this case, it turns out that the network is purely inductive
with $r=0$. Notice that a realistic network would never reach the
limit of the zero resistance. In the generic situations, the value
of the $a_{ij}$ varies from $+90^{0}$ to $-90^{0}$. Further, the
phase $a_{ij}=-90^{0}$ is also not feasible in the real situation,
as the network possesses a finite capacitance. Thus, the phases
for the inductor and capacitor circuits can be defined as
\begin{scriptsize}
\begin{eqnarray}
a_{(1)ij}&=& tan^{-1}(\frac{X_{L_{ij}}}{r_{ij}}), \ a_{ij}=-90^{0}
\nonumber \\ a_{(2)ij}&=& tan^{-1} (\frac{X_{C_{ij}}}{r_{ij}}),\
a_{ij}= 90^{0}
\end{eqnarray}
\end{scriptsize}

For a general consideration of the network fluctuations, we have
non-zero values for the network parameters, viz., $r$, $L$ and
$C$, and thus the general phase angle $a_{(3)ij}$ is defined as
\begin{scriptsize}
\begin{eqnarray}
a_{(3)ij}= tan^{-1}( \frac{X_{L_{ij}}- X_{C_{ij}}}{r_ij})
\end{eqnarray}
\end{scriptsize}

We take an account of the fact that the efficiency of the power
flow on a transmission line is analyzed by the phases of the
impedance pertaining to the transmission lines. As per the
consideration of the next subsection, our method offers a
non-linear characterization for the generic component of a
realistic network, which we suppose neither a purely inductive nor
capacitive component.

\subsection{Hypothesis of the Intrinsic Geometry}
In this subsection, we recall the motivation for intrinsic
geometric analysis and set up the notations for the subsequent
computations. Following the notations of the previous subsection,
a given network can reach a local equilibrium, if we can fix one
of the phases of the power network. The logic simply follows from
the fact that the sum of the three angles of the trigone is a
constant. To be specific, let us illustrate the consideration of
the intrinsic geometry for the case of two parameter
configurations. To be concrete, let the parameters be $\{ a_1,a_2
\}$ and let $S(a_1,a_2)$ be a smooth function of the network
(real, imaginary ) phases as defined in the
Eqns.(\ref{gerenalrealpower}, \ref{generalimaginarypower}) or any
of their real combinations. For a given $S(a_1,a_2)$, the
components of the correlation functions are described as the
Hessian matrix $Hess(S(a_1,a_2))$ of the generalized power
function under the flow of the parameters. Following this
consideration, the components of the intrinsic metric tensor are
given by
\begin{eqnarray}
g_{a_1a_1}= \frac{\partial^2 S}{\partial a_1^2},\ \ 
g_{a_1a_2}= \frac{\partial^2 S}{{\partial a_1}{\partial a_2}},\ \ 
g_{a_2a_2}= \frac{\partial^2 S}{\partial a_2^2}
\end{eqnarray}
The components of the intrinsic metric tensor are associated to
the respective pair correlation functions of the concerned power
flow. It is worth mentioning that the co-ordinates of the
underlying power factor lie on the surface of the parameters,
which in the statistical sense, gives the origin of the
fluctuations in the network. This is because the components of the
metric tensor comprise the Gaussian fluctuations of the network
power, which is a function of the parameters of the power
configuration. For a given network, the local stability of the
underlying system requires both the principle components to be
positive. In this concern, the diagonal components of the metric
tensor, $\{ { g_{a_ia_i} \mid i \in {1,2}}\}$ signify the heat
capacities of the system, and thus they are required to remain
positive definite quantities
\begin{eqnarray}
g_{a_ia_i} &>& 0, \ i= \ 1,2
\end{eqnarray}
From the perspective of intrinsic geometry, the stability
properties of the network flows can thus be divulged from the
positivity of the determinant of the metric tensor. For the
Gaussian fluctuations of the two charge equilibrium power
configurations, the existence of a positive definite volume form
on the power surface imposes such a stability condition.
Specifically, a power supplying configuration is said to be stable
if the determinant of the tensor
\begin{eqnarray}
\Vert g \Vert &= &S_{a_1a_1}S_{a_2a_2}- S_{a_1a_2}^2
\end{eqnarray}
remains positive. For the two parameters networks, the geometric
quantities corresponding to the chosen power elucidate the typical
features of the Gaussian fluctuations about an ensemble of
equilibrium states. Subsequently, the intrinsic scalar curvature,
as a global invariant, accompanies the information of the
correlation volume of the underlying power fluctuations.
Explicitly, the scalar curvature $ R $ takes the following form:
\begin{eqnarray}
R&=& -\frac{1}{2}(S_{a_1a_1}S_{a_2 a_2}- S_{a_1 a_2}^2)^{-2}
(S_{a_2 a_2}S_{a_1a_1a_1}S_{a_1 a_2 a_2} \nonumber \\ && +
S_{a_1a_2}S_{a_1a_1a_2}S_{a_1a_2a_2}+S_{a_1a_1}S_{a_1a_1a_2}
S_{a_2a_2a_2} \nonumber \\ &&
+S_{a_1a_2}S_{a_1a_1a_1}S_{a_2a_2a_2}-
S_{a_1a_1}S_{a_1a_2a_2}^2-S_{a_2a_2}S_{a_1a_1a_2}^2)
\end{eqnarray}
Notice that the zero scalar curvature indicates that the power of
the network fluctuates independently of the phases, while a
divergent scalar curvature signifies a sort of phase transition,
indicating an ensemble of highly correlated pixels of information
on the power surface. In the case of black hole physics, Ruppeiner
has interpreted the assumption ``that all the statistical degrees
of freedom of a black hole live on the black hole event horizon''
as an indication that the state-space scalar curvature signifies
the average number of correlated Planck areas on the event horizon
of the black hole \cite{RuppeinerRMP}. For the case of the two
parameter systems, the above analysis of the surface shows that
the scalar curvature and curvature tensor are related by
\begin{eqnarray}
R(a_1,a_2)=\frac{2}{\Vert g \Vert}R_{a_1a_2a_1a_2}
\end{eqnarray}

The scalar curvature thus defined offers the nature of the long
range global correlation and underlying phase transitions
originating from the power flow. In this sense, we anticipate that
an ensemble of signals corresponding to the network are
statistically interacting, if the underlying power configuration
has a non-zero scalar curvature. Incrementally, we may notice
further that the configurations under present consideration are
allowed to be effectively attractive or repulsive, and weakly
interacting, in general. The intrinsic geometric analysis further
provides a set of physical indications encoded in the
geometrically invariant quantities, e.g., scalar curvature and
other geometrically non-trivial objects. For the electrical
network, the underlying analysis would involve an ensemble or
subensemble of the equilibrium configuration forming a statistical
basis about the Gaussian distribution. With this brief
introduction, we shall now proceed to systematically analyze the
underlying stability structures of the network fluctuations in the
real, imaginary power flows and their joint effects on the
network.

\section{Real Power Flow}
Let us first describe the intrinsic stability of the electrical
network with a given power factor. Following the
Eqn.(\ref{gerenalrealpower}), the power defined with a set of
desired corrections over the network power factors, chosen as the
network variables $a_1, a_2$ for the present analysis, is given by
\begin{scriptsize}
\begin{eqnarray} \label{realpower}
 P(a_1,a_2):= \frac{V^2}{R_0(1+(tan(a_1)-tan(a_2))^2)}
\end{eqnarray}
\end{scriptsize}
The components of the correlation functions are described as the
Hessian matrix $Hess(P(a_1,a_2))$ of the concerned power under the
tuning response function. Following the Eqn.(\ref{realpower}), the
components of the metric tensor are
\begin{scriptsize}
\begin{eqnarray}
g_{a_1a_1} &=& \frac{2c_2^3V^2}{R_0} \frac{n^R_{11}}{r^R_{11}} \nonumber \\
g_{a_1a_2} &=& -\frac{2 c_2^2 c_1^2 V^2}{R_0}
\frac{n^R_{12}}{r^R_{12}} \nonumber \\
g_{a_2a_2} &=& \frac{2 c_1^3 V^2}{R_0} \frac{n^R_{22}}{r^R_{22}}
\end{eqnarray}
\end{scriptsize}
In this framework, we find that the geometric nature of the
parametric pair correlations offers the notion of fluctuating
networks. Thus, the fluctuating parameters may be easily divulged
in terms of the intrinsic parameters of the underlying network
configurations. For a given network, it is evident that the
principle components of the metric tensor signify self pair
correlations, which are positive definite functions over a range
of the parameters. In order to simplify the subsequent notations,
let us define the cosine and sine variables as
\begin{scriptsize}
\begin{eqnarray}
cos(a_i)&:=&c_i, i=1,2 \nonumber \\
 sin(a_i)&:=& s_i, i=1,2
\end{eqnarray}
\end{scriptsize}
We have arrived at the conclusion that the numerator of the local
pair correlation functions are expressed as the following
trigonometric polynomials:
\begin{scriptsize}
\begin{eqnarray}
n^R_{11}&:=& -6 c_1^4 c_2+6 s_1 c_2^2 s_2 c_1^3 -c_2^3+6 c_1^4
c_2^3 \nonumber \\ && -2 s_1 s_2 c_1^3 +3 c_1^2 c_2-c_1^2 c_2^3 \nonumber \\
n^R_{12}&:=& 6 s_1 c_2 s_2 c_1-3 c_1^2 +7 c_1^2 c_2^2 -3 c_2^2   \nonumber \\
n^R_{22}&:=&-c_1^3 c_2^2-c_1^3-2 s_1 c_2^3 s_2 +6 c_2^3 s_1 s_2
c_1^2+6 c_1^3 c_2^4  \nonumber \\ && -6 c_1 c_2^4+3 c_1 c_2^2
\end{eqnarray}
\end{scriptsize}
We further notice a similar conclusion for the denominator of the
local pair correlation functions. In  general, we find, for the
real power flow pair correlations, that the denominator of the
local pair correlation functions are
\begin{scriptsize}
\begin{eqnarray}
r^R_{11}&:=& -c_1^6-15 c_1^2 c_2^4+42 c_1^4 c_2^4  -27 c_1^4
c_2^6+15 c_1^6 c_2^2-27 c_1^6 c_2^4 \nonumber \\&& +15 c_1^2 c_2^6
+6 s_1 c_2 s_2 c_1^5 +20 s_1 c_2^3 s_2 c_1^3 -15 c_1^4 c_2^2
\nonumber \\&& +6 s_1 c_2^5 s_2 c_1-c_2^6  +13 c_1^6 c_2^6 -20 s_1
c_2^3 s_2 c_1^5 \nonumber \\&& -20 s_1 c_2^5
s_2 c_1^3 +14 c_1^5 c_2^5 s_1 s_2  \nonumber \\
r^R_{12}&:=& -c_1^6-15 c_1^2 c_2^4+42 c_1^4 c_2^4  -27 c_1^4 c_2^6
+15 c_1^6 c_2^2-27 c_1^6 c_2^4 \nonumber \\&& +15 c_1^2 c_2^6 +6
s_1 c_2 s_2 c_1^5  +20 s_1 c_2^3 s_2 c_1^3 -15 c_1^4 c_2^2
\nonumber \\&& +6 s_1 c_2^5 s_2 c_1-c_2^6  +13 c_1^6 c_2^6-20 s_1
c_2^3 s_2 c_1^5 \nonumber \\&& -20 s_1 c_2^5
s_2 c_1^3 +14 c_1^5 c_2^5 s_1 s_2 \nonumber \\
r^R_{22}&:=& -c_1^6-15 c_1^2 c_2^4+42 c_1^4 c_2^4 -27 c_1^4
c_2^6+15 c_1^6 c_2^2-27 c_1^6 c_2^4 \nonumber \\&& +15 c_1^2
c_2^6+6 s_1 c_2 s_2 c_1^5 +20 s_1 c_2^3 s_2 c_1^3 -15 c_1^4 c_2^2
\nonumber \\&& +6 s_1 c_2^5 s_2 c_1-c_2^6+13 c_1^6 c_2^6-20 s_1
c_2^3 s_2 c_1^5 \nonumber \\&&-20 s_1 c_2^5 s_2 c_1^3 +14 c_1^5
c_2^5 s_1 s_2
\end{eqnarray}
\end{scriptsize}
It is worth mentioning that the real network is well-behaved for
the generic values of the parameters. Over the domain of the $\{
a_1, a_2 \}$, we observe that the Gaussian fluctuations form a set
of stable correlations, if the determinant of the metric tensor
\begin{scriptsize}
\begin{eqnarray} \label{deteqR}
g = \frac{8 V^4 c_2^3 c_1^3}{R_0^2} \frac{n^R_g}{r^R_g}
\end{eqnarray}
\end{scriptsize}
remains a positive function on the power factor surface
$(M_2(R),g)$. In terms of the trigonometric polynomial, we obtain
that the numerator of the determinant of the metric tensor can be
expressed as
\begin{scriptsize}
\begin{eqnarray}
n^R_g:&=& 12 s_1 s_2 c_1^4 c_2^4-4 c_1^3 c_2^3 +4 c_1^3 c_2-c_1^5
c_2 -c_1 c_2^5 \nonumber
\\&& +4 c_1 c_2^3 -c_1^4 s_1 s_2 -c_1^4 s_1 s_2 c_2^2 -c_2^4 s_1 s_2
c_1^2 \nonumber \\&& -c_2^4 s_1 s_2 -7 c_1^3 c_2^5  -6 c_2^2 c_1^2
s_1 s_2 -7 c_1^5 c_2^3 \nonumber
\\&& +12 c_1^5 c_2^5
\end{eqnarray}
\end{scriptsize}
As per the expectation, the denominator of the determinant of the
metric tensor turns out to be given by the following trigonometric
polynomial:
\begin{scriptsize}
\begin{eqnarray}
r^R_g:&=&-210 c_1^4 c_2^6-210 c_1^6 c_2^4  +910 c_1^6 c_2^6
-c_1^{10} -c_2^{10} +121 c_2^{10} c_1^{10} \nonumber \\&& +122
c_2^9 c_1^9 s_1 s_2 +252 c_1^5 c_2^5 s_1 s_2 -584 s_1 c_2^7 s_2
c_1^5 +332 s_1 c_2^9 s_2 c_1^5 \nonumber \\&& +808 s_1 c_2^7 s_2
c_1^7 +120 s_1 c_2^3 s_2 c_1^7 -120 c_2^3 c_1^9 s_1 s_2 +10 s_1
s_2 c_2 c_1^9 \nonumber \\&& +45 c_2^{10} c_1^2 -250 c_2^{10}
c_1^4 +490 c_2^{10} c_1^6 -405 c_2^10 c_1^8 +45 c_2^2 c_1^{10}
\nonumber \\&& -120 s_1 c_2^9 s_2 c_1^3+10 s_1 c_2^9 s_2 c_1 -584
s_1 c_2^5 s_2 c_1^7 -344 s_1 c_2^9 s_2 c_1^7 \nonumber \\&& -344
s_1 c_2^7 s_2 c_1^9 +332 c_2^5 c_1^9 s_1 s_2+120 s_1 c_2^7 s_2
c_1^3 -250 c_2^4 c_1^{10} \nonumber \\&& +460 c_2^4 c_1^8-45 c_1^2
c_2^8 +490 c_2^6 c_1^{10}-1190 c_2^8 c_1^6 +1180 c_2^8 c_1^8
\nonumber \\&& -405 c_2^8 c_1^10 -1190 c_2^6 c_1^8-45 c_2^2 c_1^8
+460 c_2^8 c_1^4
\end{eqnarray}
\end{scriptsize}
Herewith, the behavior of the determinant of the metric tensor
shows that such a real power flow becomes unstable for the
specific values of the parameters. For generic $R_0$ and $V$, the
nature of the determinant of the metric tensor is depicted in the
Eqn.(\ref{deteqR}). It is worth mentioning further that the
electric networks become unstable in the limit of vanishing
$\{a_1, a_2\}$.
\begin{figure}
\hspace*{0.5cm}
\includegraphics[width=8.0cm,angle=-90]{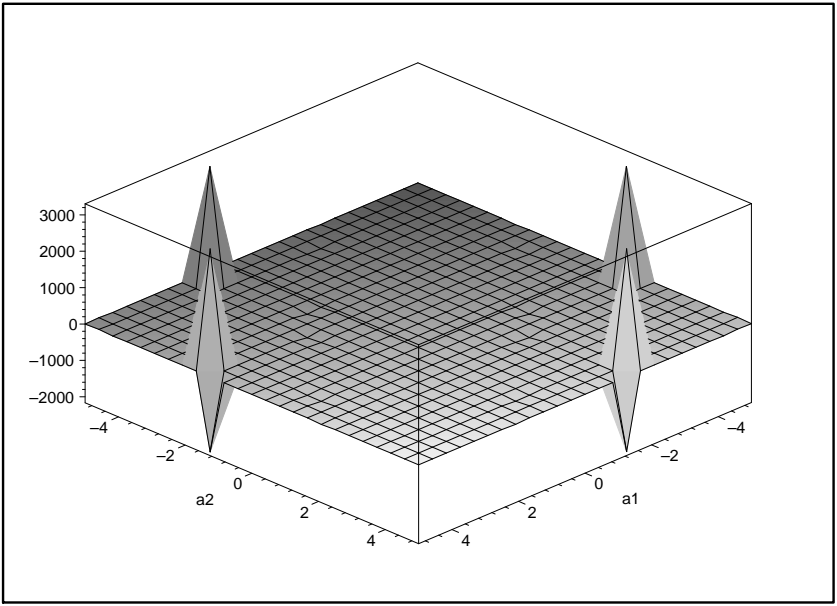}
\caption{The determinant of the metric tensor plotted as the
function of the power factors $a_1, a_2$, describing the real
power fluctuations in electrical networks.}  \label{det3d}
\vspace*{0.5cm}
\end{figure}

\begin{figure}
\hspace*{0.5cm}
\includegraphics[width=8.0cm,angle=-90]{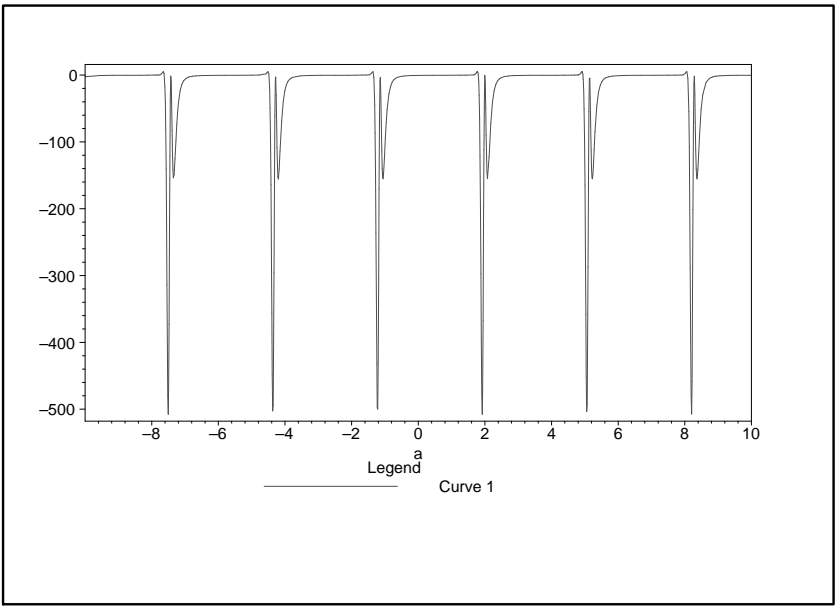}
\caption{The determinant of the metric tensor plotted as the
function of the equal power factor $a:=a_1= a_2$, describing the
real power fluctuations in electrical networks.}   \label{det2d}
\vspace*{0.5cm}
\end{figure}
In order to explain the nature of transformation of the
$\{a_1,a_2\}$ forming the intrinsic surface, let us explore the
functional behavior of the associated scalar curvature. Our
computation shows that the scalar curvature reduces to the
following form:
\begin{scriptsize}
\begin{eqnarray}
R = \frac{1}{4R_0c_1^2c_2^2V^2} \frac{n^R_R}{r^R_R}
\end{eqnarray}
\end{scriptsize}
The numerator and denominator of the scalar curvature take the
following trigonometric polynomial expression:
\begin{scriptsize}
\begin{eqnarray}
n^R_R&:=& -42 c_1^4 c_2^6-42 c_1^6 c_2^4+444 c_1^6 c_2^6  +3
c_1^{10}+3 c_2^{10} +702 c_2^{10} c_1^{10} \nonumber \\&& +702
c_2^9 c_1^9 s_1 s_2 +84 c_1^5 c_2^5 s_1 s_2 -280 s_1 c_2^7 s_2
c_1^5 +356 s_1 c_2^9 s_2 c_1^5 \nonumber \\&& +1080 s_1 c_2^7 s_2
c_1^7 -24 s_1 c_2^3 s_2 c_1^7 +24 c_2^3 c_1^9 s_1 s_2 -18 s_1 s_2
c_2 c_1^9 \nonumber \\&& -38 c_2^{10} c_1^2-70 c_2^{10} c_1^4 +752
c_2^{10} c_1^6 -1349 c_2^{10} c_1^8-38 c_2^2 c_1^10  \nonumber
\\&& +24 s_1 c_2^9 s_2 c_1^3 -18 s_1 c_2^9 s_2 c_1 -280 s_1 c_2^5
s_2 c_1^7 -1000 s_1 c_2^9 s_2 c_1^7  \nonumber \\&& -1000 s_1
c_2^7 s_2 c_1^9 +356 c_2^5 c_1^9 s_1 s_2 -24 s_1 c_2^7 s_2 c_1^3
-70 c_2^4 c_1^{10} \nonumber \\&& +72 c_2^4 c_1^8 +39 c_1^2 c_2^8
+752 c_2^6 c_1^10-1082 c_2^8 c_1^6 +2288 c_2^8 c_1^8  \nonumber
\\&& -1349 c_2^8 c_1^{10}-1082 c_2^6 c_1^8 +39 c_2^2 c_1^8
+72 c_2^8 c_1^4
\end{eqnarray}
\end{scriptsize}
The denominator of the scalar curvature can be expressed as the
trigonometric polynomial
\begin{scriptsize}
\begin{eqnarray}
r^R_R&:=&-c_1^6+6 s_1 c_2 s_2 c_1^5+20 s_1 c_2^3 s_2 c_1^3 +6 s_1
c_2^5 s_2 c_1+8 s_1 c_2^3 s_2 c_1^5 \nonumber \\&& +8 s_1 c_2^5
s_2 c_1^3-15 c_1^2 c_2^4+6 c_1^4 c_2^4 +53 c_1^4 c_2^6+4 c_1^6
c_2^2+53 c_1^6 c_2^4  \nonumber \\&& +4 c_1^2 c_2^6 -15 c_1^4
c_2^2-48 c_1^6 c_2^6 +c_1^8+c_2^8 -70 c_1^5 c_2^5 s_1 s_2-c_2^6
\nonumber \\&& -12 s_1 c_2^7 s_2 c_1^5 +72 s_1 c_2^7 s_2 c_1^7 -12
s_1 c_2^3 s_2 c_1^7  -12 s_1 c_2^5 s_2 c_1^7  \nonumber \\&& -12
s_1 c_2^7 s_2 c_1^3 -11 c_2^4 c_1^8+2 c_1^2 c_2^8 -48 c_2^8 c_1^6
+72 c_2^8 c_1^8 \nonumber \\&& -48 c_2^6 c_1^8 +2 c_2^2 c_1^8 -11
c_2^8 c_1^4
\end{eqnarray}
\end{scriptsize}

\begin{figure}
\hspace*{0.5cm}
\includegraphics[width=8.0cm,angle=-90]{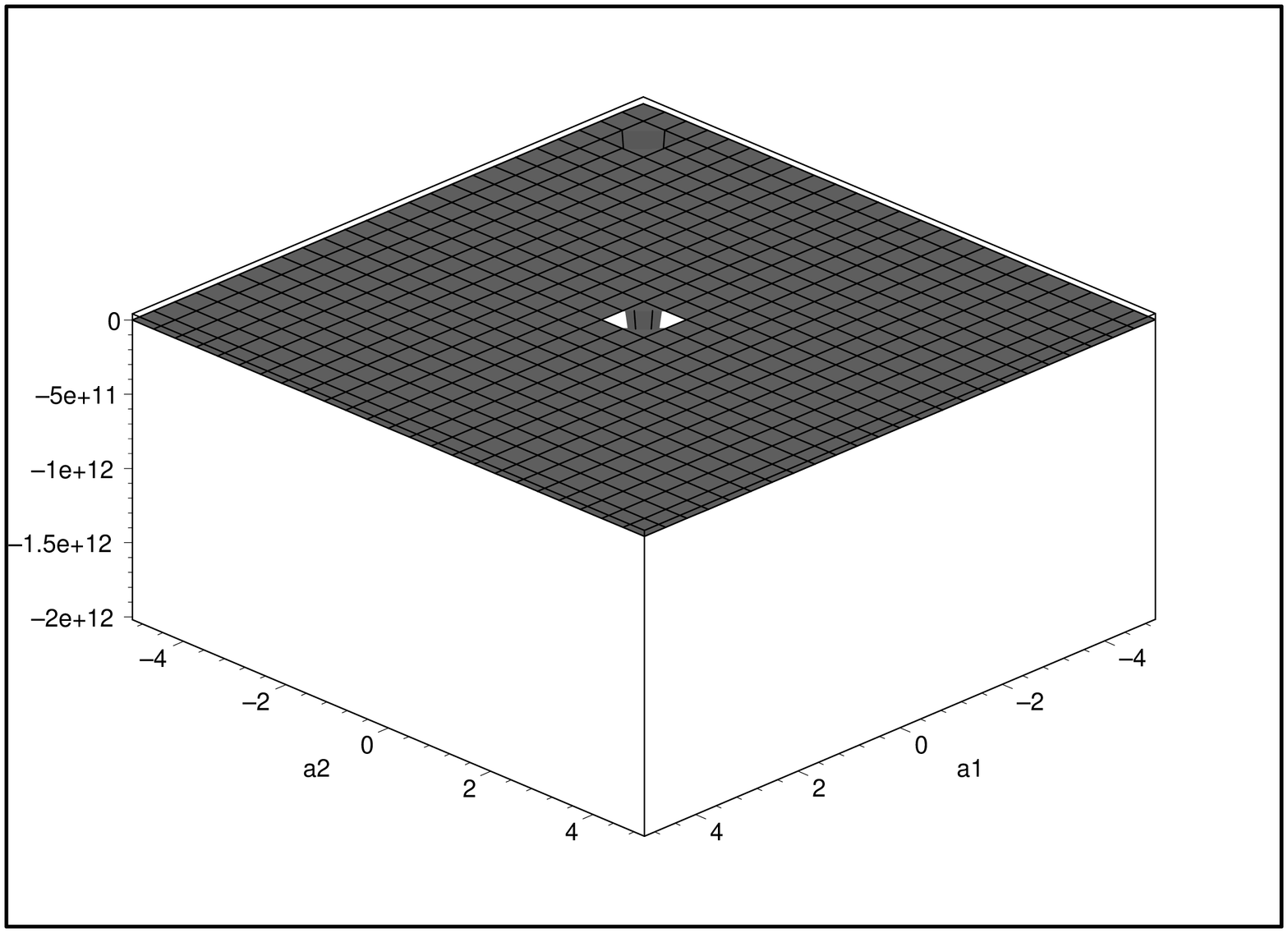}
\caption{The curvature scalar plotted as a function of the power
factors $a_1, a_2$, describing the real power fluctuations in
electrical networks.} \label{cur3d} \vspace*{0.5cm}
\end{figure}

\begin{figure}
\hspace*{0.5cm}
\includegraphics[width=8.0cm,angle=-90]{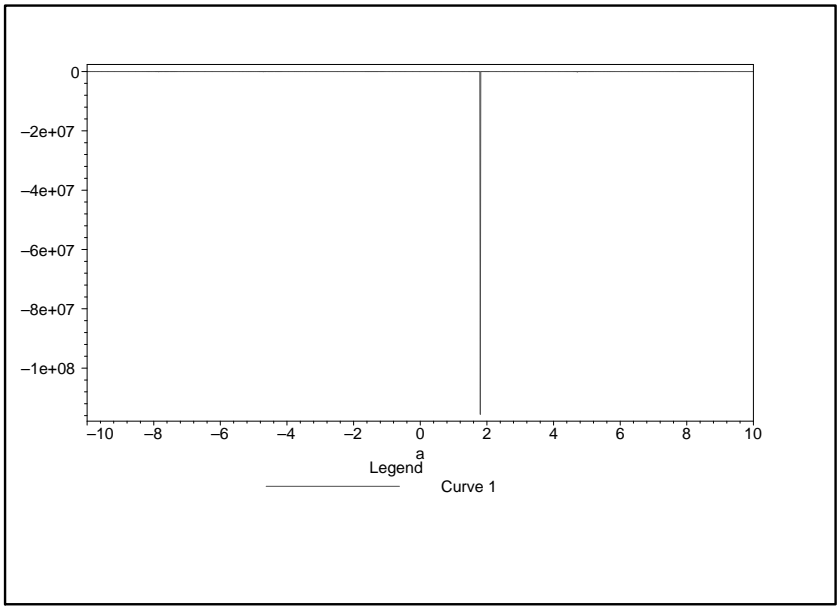}
\caption{The curvature scalar plotted as a function of the equal
power factor $a:= a_1= a_2$, describing the real fluctuations in
electrical networks.} \label{cur2d} \vspace*{0.5cm}
\end{figure}

We find that a typical real power network is globally correlated
over all the generic Gaussian fluctuations of the parameters
$\{a_1,a_2\}$, unless $n^R_R=0$. For the above real power
networks, we observe that the scalar curvature diverges in the
limit $r^R_R=0$, showing a signature of the global instability on
the $\{a_1,a_2\}$ surface. Thus, the intrinsic geometric analysis
shows that a real power network is interacting and locally stable
over the surface of fluctuation, if the network parameters
$\{a_1,a_2\}$ are properly chosen.

For the choices $V=1$ and $R_0=1$, the Fig.(\ref{det3d}) shows the
determinant of the metric tensor. These plots explicate the nature
of the stability in the real power flow in the power networks. The
corresponding plot for the scalar curvature is depicted in the
Fig.(\ref{cur3d}). This plot shows the global nature of real power
flow in the electrical network under the effects of Gaussian
fluctuations of the parameters.

For the equal phases, viz., $a_1=a$ and $a_2=a$, the surface plots
of the determinant of the metric tensor and scalar curvature are
respectively shown in the Figs.(\ref{det2d}) and (\ref{cur2d}). We
observe that the stability of the real power networks exists in
certain bands. This follows from the fact that the instability is
present only for a set of specific equal values of the parameters.
For a limiting equal phases network, the limiting scalar curvature
interestingly simplifies to the shown shape. In particular, the
corresponding peaks in the Figs.(\ref{cur3d}) and (\ref{cur2d}) of
the scalar curvature indicate the graphical nature of the global
instability in the real power networks. Physically, the peaks in
the curvature show the presence of non-trivial interactions in the
network.

\section{Imaginary Power Flow}
In the present section, we analyze the nature of an ensemble of
fluctuating electrical networks generated by a pair $a_1, a_2$. To
focus on the most general case, we chose the variable $a_1, a_2$
as a function of $L$, $C$ and $r$ of the chosen network. Following
Eqn.(\ref{generalimaginarypower}), when the imaginary power
\begin{scriptsize}
\begin{eqnarray} \label{imaginarypower}
Q(a_1,a_2):=
\frac{V^2}{R_0}\frac{(tan(a_1)-tan(a_2))}{(1+(tan(a_1)-tan(a_2))^2)}
\end{eqnarray}
\end{scriptsize}
is allowed to fluctuate as a function of the $a_1, a_2$, we may
again exploit the definition of the Hessian function
$Hess(Q(a_1,a_2))$ of the imaginary power. Herewith, the
components of the metric tensor are given by
\begin{scriptsize}
\begin{eqnarray}
g_{a_1a_1} &=& -\frac{2c_2^2V^2}{R_0} \frac{n^I_{11}}{r^I_{11}} \nonumber \\
g_{a_1a_2} &=& \frac{2 c_2^2 c_1^2 V^2}{R_0}
\frac{n^I_{12}}{r^I_{12}} \nonumber \\
g_{a_2a_2} &=& -\frac{2 c_1^2 V^2}{R_0} \frac{n^I_{22}}{r^I_{22}}
\end{eqnarray}
\end{scriptsize}
In this case, the numerators of the local pair correlation
functions are expressed as the following trigonometric
polynomials:
\begin{scriptsize}
\begin{eqnarray}
n^I_{11}&:=& -5 c_2^3 c_1^2 s_2-s_1 c_1^3 +8 c_2^2 s_1 c_1^3 -4
c_2 s_2 c_1^4 +3 c_2 s_2 c_1^2 \nonumber \\&& -3 c_2^2 s_1 c_1 +8
c_1^4 s_2 c_2^3-7 s_1 c_1^3
c_2^4 +s_2 c_2^3 +c_2^4 s_1 c_1 \nonumber \\
n^I_{12}&:=& c_2^3 s_1+3 c_1^2 s_1 c_2-c_1^3 s_2 -7 c_2^3 c_1^2
s_1+ 7c_2^2 s_2 c_1^3 -3 c_2^2 s_2 c_1 \nonumber \\
n^I_{22}&:=& -8 c_2^3 c_1^2 s_2-s_1 c_1^3 +5 c_2^2 s_1 c_1^3 -c_2
s_2 c_1^4 +3 c_2 s_2 c_1^2 \nonumber \\&& -3 c_2^2 s_1 c_1 +7
c_1^4 s_2 c_2^3-8 s_1 c_1^3 c_2^4 +s_2 c_2^3 +4 c_2^4 s_1 c_1
\end{eqnarray}
\end{scriptsize}
While, the denominators of the local pair correlation functions
take the following trigonometric expressions:
\begin{scriptsize}
\begin{eqnarray}
r^I_{11}&:=& 13 c_1^6 c_2^6+14 c_1^5 c_2^5 s_1 s_2 +6 s_1 c_2^5
s_2 c_1+15 c_1^2 c_2^6 \nonumber
\\&& +6 s_1 c_2 s_2 c_1^5-20 c_1^3 c_2^5 s_1 s_2
-20 c_1^5 c_2^3 s_1 s_2-15 c_1^2 c_2^4 \nonumber
\\&& +20 c_1^3 c_2^3 s_1 s_2-27 c_1^4 c_2^6
+15 c_1^6 c_2^2-27 c_1^6 c_2^4+42 c_1^4 c_2^4 \nonumber \\&&
-15 c_1^4 c_2^2 -c_2^6-c_1^6  \nonumber \\
r^I_{12}&:=& 13 c_1^6 c_2^6+14 c_1^5 c_2^5 s_1 s_2 +6 s_1 c_2^5
s_2 c_1+15 c_1^2 c_2^6 \nonumber
\\&& +6 s_1 c_2 s_2 c_1^5-20 c_1^3 c_2^5 s_1 s_2
-20 c_1^5 c_2^3 s_1 s_2-15 c_1^2 c_2^4 \nonumber
\\&& +20 c_1^3 c_2^3 s_1 s_2-27 c_1^4 c_2^6
+15 c_1^6 c_2^2-27 c_1^6 c_2^4+42 c_1^4 c_2^4 \nonumber \\&&
-15 c_1^4 c_2^2-c_2^6-c_1^6 \nonumber \\
r^I_{22}&:=& 13 c_1^6 c_2^6+14 c_1^5 c_2^5 s_1 s_2 +6 s_1 c_2^5
s_2 c_1+15 c_1^2 c_2^6 \nonumber
\\&& +6 s_1 c_2 s_2 c_1^5-20 c_1^3 c_2^5 s_1 s_2
-20 c_1^5 c_2^3 s_1 s_2-15 c_1^2 c_2^4 \nonumber
\\&& +20 c_1^3 c_2^3 s_1 s_2-27 c_1^4 c_2^6
+15 c_1^6 c_2^2-27 c_1^6 c_2^4+42 c_1^4 c_2^4 \nonumber \\&& -15
c_1^4 c_2^2-c_2^6-c_1^6
\end{eqnarray}
\end{scriptsize}
It follows that the pure pair correlations $\{ g_{11}, g_{22} \}$
between the parameters $\{a_1, a_2\}$ remain positive, which is
the same as for the flow of the real power. A straightforward
computation further demonstrates the over-all nature of the
parametric fluctuations. In fact, we find that the determinant of
the metric tensor reduces to the following expression:
\begin{scriptsize}
\begin{eqnarray}
g = -\frac{V^4 c_2^3 c_1^3}{R_0^2} \frac{n^I_g}{r^I_g}
\end{eqnarray}
\end{scriptsize}
In this case, the numerator of the determinant of the metric
tensor is given by the following trigonometric polynomial:
\begin{scriptsize}
\begin{eqnarray}
n^I_g:&=& 10 c_1^6 c_2^2+37 c_1^5 c_2^5 s_1 s_2 -3 c_1^4 c_2^2-3
c_1^2 c_2^4-39 c_1^6 c_2^4 +28 c_1^4 c_2^4 \nonumber \\&& +10
c_1^2 c_2^6-39 c_1^4 c_2^6 +3 s_1 c_2^5 s_2 c_1-20 c_1^3 c_2^5 s_1
s_2 -20 c_1^5 c_2^3 s_1 s_2 \nonumber \\&& +3 s_1 c_2 s_2 c_1^5 +2
c_1^3 c_2^3 s_1 s_2+38 c_1^6 c_2^6 -c_2^6-c_1^6
\end{eqnarray}
\end{scriptsize}
Explicitly, we find that the denominator of the determinant of the
metric tensor can be presented as
\begin{scriptsize}
\begin{eqnarray}
r^I_g:&=& 252 c_1^5 c_2^5 s_1 s_2+910 c_1^6 c_2^6 -210 c_1^4 c_2^6
-210 c_1^6 c_2^4 \nonumber \\&& -c_1^{10}-c_2^{10} +122 c_1^9
c_2^9 s_1 s_2 +121 c_1^{10} c_2^{10} +10 s_1 c_2^9 s_2 c_1
\nonumber \\&& -120 s_1 c_2^9 s_2 c_1^3 -120 c_1^9 c_2^3 s_1 s_2
+332 c_1^5 c_2^9 s_1 s_2 +332 c_1^9 c_2^5 s_1 s_2 \nonumber
\\&& -344 c_1^9 c_2^7 s_1 s_2 -344 c_1^7 c_2^9 s_1 s_2
+490 c_1^6 c_2^{10} -405 c_1^8 c_2^{10}-250 c_1^{10} c_2^4
\nonumber \\&& +490 c_1^{10} c_2^6 +1180 c_2^8 c_1^8-1190 c_2^6
c_1^8 -1190 c_2^8 c_1^6 -250 c_1^4 c_2^{10}-405 c_1^{10} c_2^8
\nonumber \\&& -45 c_1^8 c_2^2 -45 c_2^8 c_1^2+460 c_1^8 c_2^4
+460 c_2^8 c_1^4 +45 c_1^{10} c_2^2+45 c_2^{10} c_1^2 \nonumber
\\&& +120 c_1^7 c_2^3 s_1 s_2 -584 c_1^7 c_2^5 s_1 s_2
+120 c_1^3 c_2^7 s_1 s_2 -584 c_1^5 c_2^7 s_1 s_2 \nonumber
\\&& +10 s_1 s_2 c_1^9 c_2 +808 c_1^7 c_2^7 s_1 s_2
\end{eqnarray}
\end{scriptsize}

\begin{figure}
\hspace*{0.5cm}
\includegraphics[width=8.0cm,angle=-90]{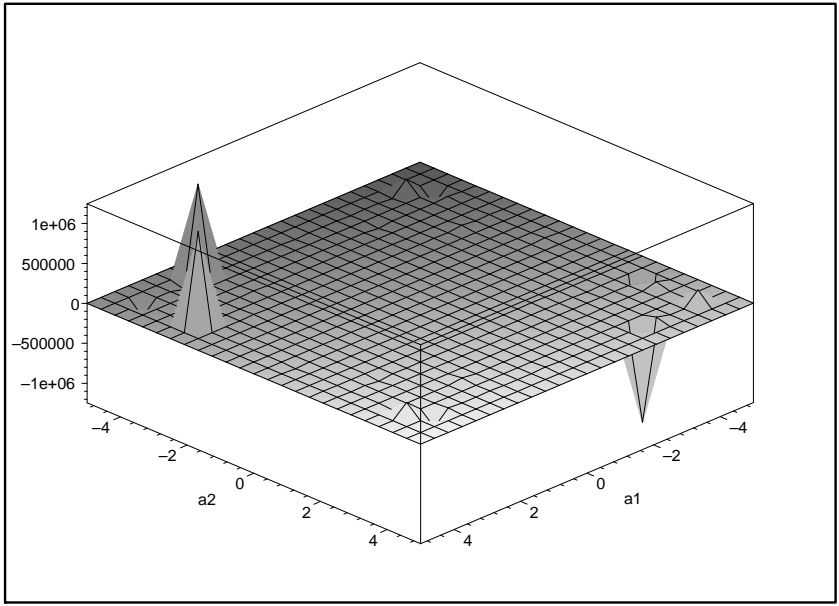}
\caption{The determinant of the metric tensor plotted as the
function of the power factors $a_1, a_2$, describing the imaginary
power fluctuations in electrical networks.}  \label{det3dIm}
\vspace*{0.5cm}
\end{figure}

\begin{figure}
\hspace*{0.5cm}
\includegraphics[width=8.0cm,angle=-90]{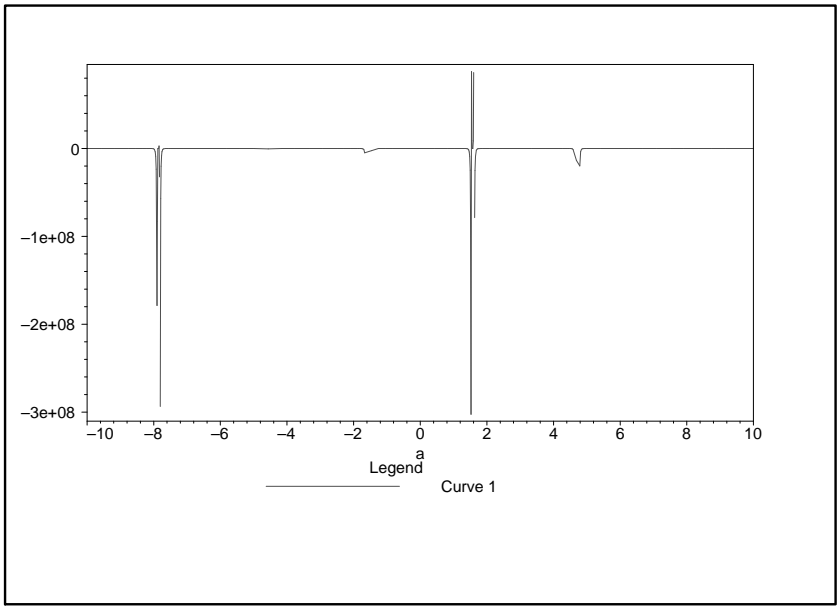}
\caption{The curvature scalar plotted as a function of the power
factors $a_1, a_2$,, describing the imaginary power fluctuations
in electrical networks.}   \label{det2dIm} \vspace*{0.5cm}
\end{figure}
It is not difficult to compute the exact expression for the scalar
curvature describing the global parametric intrinsic correlations.
In particular, we find that the scalar curvature reduces to the
following form:
\begin{scriptsize}
\begin{eqnarray}
R = -\frac{R_0}{2c_1^2c_2^2V^2}
\frac{n^{(1)I}_R+n^{(2)I}_R}{r^I_R}
\end{eqnarray}
\end{scriptsize}
In the above equation, the numerator of the scalar curvature takes
the trigonometric polynomial expression
\begin{scriptsize}
\begin{eqnarray}
n^{(1)I}_R&:=& -297 s_1 c_1^5 c_2^8+2235 s_1 c_1^5 c_2^{10} +4770
s_1 c_1^7 c_2^8-15596 s_1 c_1^7 c_2^{10} \nonumber \\&& -19538 s_1
c_1^9 c_2^8+45624 s_1 c_1^9 c_2^{10} -7857 s_1 c_1^{11} c_2^6-3954
s_1 c_1^5 c_2^{12}  \nonumber \\&& +19890 s_1 c_1^7 c_2^{12}-49482
s_1 c_1^9 c_2^{12} -947 s_2 c_1^4 c_2^11+59571 s_1 c_1^{11}
c_2^{12} \nonumber
\\&& +121 s_2 c_1^{10} c_2^3-2235 s_2 c_1^{10} c_2^5
+30498 s_1 c_1^{11} c_2^8-3653 s_2 c_1^6 c_2^9 \nonumber
\\&& -11 s_2 c_1^{14} c_2+15596 s_2 c_1^{10}
c_2^7 -45624 s_2 c_1^{10} c_2^9+60771 s_2 c_1^{10} c_2^{11}
\nonumber \\&& -355 s_2 c_1^{12} c_2^3+3954 s_2 c_1^{12} c_2^5
-19890 s_2 c_1^{12} c_2^7+49482 s_2 c_1^{12} c_2^9 \nonumber \\&&
-59571 s_2 c_1^{12} c_2^{11}-60771 s_1 c_1^{11} c_2^{10} +19538
s_2 c_1^8 c_2^9+7857 s_2 c_1^6 c_2^{11}  \nonumber \\&& -30498 s_2
c_1^8 c_2^{11}+2080 s_1 c_1^5 c_2^{14} -8902 s_1 c_1^7
c_2^{14}+20139 s_1 c_1^9 c_2^{14} \nonumber
\\&& -22407 s_1 c_1^{11} c_2^{14}+8902 s_2 c_1^{14} c_2^7
-20139 s_2 c_1^{14} c_2^9+22407 s_2 c_1^{14} c_2^{11} \nonumber \\
n^{(2)I}_R&:=& 11 s_1 c_1 c_2^{14}-2080 s_2 c_1^{14} c_2^5 -243
s_1 c_1^3 c_2^{14}-9450 s_2 c_2^{13}c_1^{14} \nonumber
\\&& +61 s_1 c_1^{13} c_2^2-812 s_1 c_1^{13}
c_2^4  +5044 s_1 c_1^{13} c_2^6-16769 s_1 c_1^{13} c_2^8 \nonumber
\\&& +30165 s_1 c_1^{13} c_2^{10}-27138 s_1 c_1^{13} c_2^{12}
+9450 s_1 c_1^{13} c_2^{14} \nonumber \\&& -5044 s_2 c_1^6
c_2^{13}-61 s_2 c_1^2 c_2^{13} +220 s_2 c_1^4 c_2^9-121 s_1 c_1^3
c_2^{10} \nonumber \\&& +947 s_1 c_1^{11} c_2^4+355 s_1 c_1^3
c_2^{12} -45 s_1 c_1^{11} c_2^2+27138 c_1^{12} c_2^{13} s_2
\nonumber \\&&  +330 s_2 c_1^6 c_2^7+3653 s_1 c_1^9 c_2^6 -220 s_1
c_1^9 c_2^4-10 s_1 c_1 c_2^{12} \nonumber \\&& +297 s_2 c_1^8
c_2^5-330 s_1 c_1^7 c_2^6 +45 s_2 c_1^2 c_2^{11}+10 s_2 c_1^{12}
c_2 \nonumber \\&&  -4770 s_2 c_1^8 c_2^7-30165 s_2 c_1^{10}
c_2^{13}+812 s_2 c_1^4 c_2^{13}-s_1 c_1^{13} \nonumber \\&& +16769
s_2 c_1^8 c_2^{13} +s_2 c_2^{13}+243 s_2 c_1^{14} c_2^3
\end{eqnarray}
\end{scriptsize}
The function $r^I_R$ appearing in the denominator of the scalar
curvature can be written as the following polynomial:
\begin{scriptsize}
\begin{eqnarray}
r^I_R&:=& 42 c_1^6 c_2^6+c_2^{12}+c_1^{12}+5344 c_1^9 c_2^9 s_1
s_2 +10447 c_1^{10} c_2^{10}+4383 c_1^{12} c_2^8 \nonumber \\&&
+100 c_1^3 c_2^{11} s_1 s_2 -708 c_1^5 c_2^{11} s_1 s_2 +100
c_1^{11} c_2^3 s_1 s_2 +2528 c_1^7 c_2^{11} s_1 s_2 \nonumber \\&&
-4406 c_1^9 c_2^{11} s_1 s_2 +2528 c_1^{11} c_2^7 s_1 s_2 -4406
c_1^{11} c_2^9 s_1 s_2 +2812 c_1^{11} c_2^{11} s_1 s_2 \nonumber
\\&& -6 c_1 c_2^{11} s_1 s_2 -708 c_1^{11} c_2^5 s_1 s_2 -6 c_1^11
c_2 s_1 s_2 -29 c_1^2 c_2^{12}-29 c_1^{12} c_2^2 \nonumber \\&&
+4383 c_1^8 c_2^{12} -5813 c_1^{10} c_2^{12}+2813 c_1^{12}
c_2^{12} -1598 c_1^6 c_2^{12} -1598 c_1^{12} c_2^6 \nonumber \\&&
-5813 c_1^{12} c_2^{10} +307 c_1^4 c_2^{12}+307 c_1^{12} c_2^4 -22
s_1 c_2^9 s_2 c_1^3 -22 c_1^9 c_2^3 s_1 s_2 \nonumber \\&& +368
c_1^5 c_2^9 s_1 s_2 +368 c_1^9 c_2^5 s_1 s_2 -2132 c_1^9 c_2^7 s_1
s_2 -2132 c_1^7 c_2^9 s_1 s_2 \nonumber \\&& +1826 c_1^6 c_2^{10}
-6442 c_1^8 c_2^{10}-257 c_1^{10} c_2^4 +1826 c_1^{10} c_2^6 +2822
c_2^8 c_1^8 \nonumber \\&& -482 c_2^6 c_1^8  -482 c_2^8 c_1^6 -257
c_1^4 c_2^{10}-6442 c_1^{10} c_2^8 +27 c_1^8 c_2^4 +27 c_2^8 c_1^4
\nonumber \\&& +15 c_1^{10} c_2^2 +15 c_2^{10} c_1^2 -36 c_1^7
c_2^5 s_1 s_2 -36 c_1^5 c_2^7 s_1 s_2 +472 c_1^7 c_2^7 s_1 s_2
\end{eqnarray}
\end{scriptsize}
Consequently, we may easily analyze the underlying conclusions for
the specific considerations of the variable power factor of the
network. As in the case of real power flow, the global nature of
scalar curvature and associated phase transitions can be thus
determined over the range of power factors describing the
important network of interest.

\begin{figure}
\hspace*{0.5cm}
\includegraphics[width=8.0cm,angle=-90]{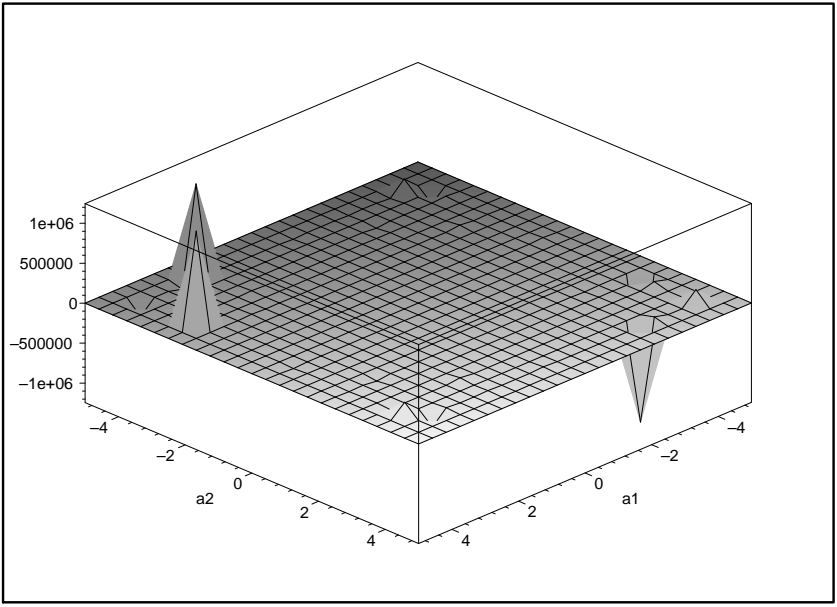}
\caption{The curvature scalar plotted as a function of the power
factors $a_1, a_2$, describing the imaginary power fluctuations in
electrical networks.}  \label{cur3dIm} \vspace*{0.5cm}
\end{figure}

As in the previous subsection, we shall focus our attention on
same electrical network and on the specific values $V=1$ and $R_0=
1$. The determinant of the metric tensor shown in the
Fig.(\ref{det3dIm}) describes the phenomenological property of the
Gaussian imaginary power fluctuations. The scalar curvature as
shown in the Fig.(\ref{cur3dIm}) depicts a couple of
antisymmetrical fluctuations for the imaginary network power flow.

For the equal values $a_1=a$ and $a_2=a$, we notice, from the
Fig.(\ref{det2dIm}), that the system acquires a couple of locally
chaotic fluctuations. Herewith, we find the surprising fact that
the imaginary power flow has a different geometric nature, and the
scalar curvature turns out to be zero in the limit of the equal
phases for the imaginary power flow. It is worth mentioning that
such a power flow is globally non-interacting.

\section{Complex Power Flow}
In order to further understand the nature of generic electrical
networks, we shall now consider the linear combination of the real
and imaginary power flows in the network. The associated joint
power flow $F$ takes the following form:
\begin{scriptsize}
\begin{eqnarray} \label{complexpower}
F(a_1,a_2):=
\frac{V^2}{R_0}\frac{(1+tan(a_1)-tan(a_2))}{(1+(tan(a_1)-tan(a_2))^2)}
\end{eqnarray}
\end{scriptsize}
In order to obtain the components of the metric tensor in the
power space, we employ the definition of the Hessian matrix as
indicated previously, and see that the metric tensor has the
following expression for the components:
\begin{scriptsize}
\begin{eqnarray}
g_{a_1a_1} &=& \frac{2c_2^2V^2}{R_0} \frac{n^C_{11}}{r^C_{11}} \nonumber \\
g_{a_1a_2} &=& -\frac{2 c_2 c_1 V^2}{R_0}
\frac{n^C_{12}}{r^C_{12}} \nonumber \\
g_{a_2a_2} &=& \frac{2 c_1^2 V^2}{R_0} \frac{n^C_{22}}{r^C_{22}}
\end{eqnarray}
\end{scriptsize}
In the above equation, the numerators of the local pair
correlation functions are expressed as the following trigonometric
polynomials:
\begin{scriptsize}
\begin{eqnarray}
n^C_{11}&:=& -c_2^3 s_2+s_1 c_1^3+3 c_1^2 c_2^2  -2 c_2 s_1 s_2
c_1^3+5 c_1^2 c_2^3 s_2 \nonumber
\\&& +4 c_1^4 c_2 s_2-c_1 c_2^4 s_1
+6 c_1^3 c_2^3 s_1 s_2-3 c_1^2 c_2 s_2 \nonumber
\\&& +7 c_1^3 c_2^4 s_1-8 c_1^4 c_2^3 s_2
+3 c_1 c_2^2 s_1+6 c_1^4 c_2^4 \nonumber
\\&& -6 c_1^4c_2^2 -8 c_1^3 c_2^2 s_1
-c_2^4 c_1^2-c_2^4 \nonumber \\
n^C_{12}&:=& 6 c_1^2 c_2^2 s_1 s_2-3 c_1^3 c_2 -3 c_1^2 s_1
c_2+c_1^3 s_2-3 c_1 c_2^3 \nonumber \\&& +7 c_1^3 c_2^3+7 s_1
c_2^3 c_1^2 -7 s_2 c_1^3 c_2^2-c_2^3 s_1 +3 c_2^2 s_2 c_1  \nonumber \\
n^C_{22}&:=& -c_2^3 s_2+s_1 c_1^3 +3 c_1^2 c_2^2 +8 c_1^2 c_2^3
s_2+c_1^4 c_2 s_2 \nonumber \\&& -4 c_1 c_2^4 s_1+6 c_1^3 c_2^3
s_1 s_2 -3 c_1^2 c_2 s_2+8 c_1^3 c_2^4 s_1 \nonumber \\&& -7 c_1^4
c_2^3 s_2+3 c_1 c_2^2 s_1 +6 c_1^4 c_2^4-c_1^4-c_1^4 c_2^2
\nonumber \\&& -5 c_1^3 c_2^2 s_1-6 c_2^4 c_1^2 -2 c_2^3 s_1 s_2
c_1
\end{eqnarray}
\end{scriptsize}
The denominators of the local pair correlation functions reduce to
the following trigonometric polynomials:
\begin{scriptsize}
\begin{eqnarray}
r^C_{11}&:=& -20 c_2^3 c_1^5 s_1 s_2-c_2^6 -c_1^6  -27 c_2^4
c_1^6+14 c_1^5 c_2^5 s_1 s_2 \nonumber \\&& +13 c_1^6 c_2^6+6 s_1
c_2^5 s_2 c_1 +15 c_1^2 c_2^6-27 c_1^4 c_2^6 \nonumber \\&& -20
c_1^3 c_2^5 s_1 s_2+6 c_2 c_1^5 s_1 s_2 +20 c_1^3 c_2^3 s_1 s_2+42
c_1^4 c_2^4 \nonumber \\&& +15 c_1^6 c_2^2-15 c_1^4 c_2^2
-15 c_2^4 c_1^2  \nonumber \\
r^C_{12}&:=& -20 c_2^3 c_1^5 s_1 s_2-c_2^6 -c_1^6 -27 c_2^4
c_1^6+14 c_1^5 c_2^5 s_1 s_2 \nonumber \\&& +13 c_1^6 c_2^6+6 s_1
c_2^5 s_2 c_1 +15 c_1^2 c_2^6-27 c_1^4 c_2^6 \nonumber \\&& -20
c_1^3 c_2^5 s_1 s_2+6 c_2 c_1^5 s_1 s_2 \nonumber \\&& +20 c_1^3
c_2^3 s_1 s_2+42 c_1^4 c_2^4 +15 c_1^6 c_2^2-15 c_1^4 c_2^2
-15 c_2^4 c_1^2 \nonumber \\
r^C_{22}&:=& -20 c_2^3 c_1^5 s_1 s_2-c_2^6 -c_1^6 -27 c_2^4
c_1^6+14 c_1^5 c_2^5 s_1 s_2 \nonumber \\&& +13 c_1^6 c_2^6+6 s_1
c_2^5 s_2 c_1 +15 c_1^2 c_2^6-27 c_1^4 c_2^6 \nonumber \\&& -20
c_1^3 c_2^5 s_1 s_2+6 c_2 c_1^5 s_1 s_2 +20 c_1^3 c_2^3 s_1 s_2
\nonumber \\&& +42 c_1^4 c_2^4 +15 c_1^6 c_2^2-15 c_1^4 c_2^2-15
c_2^4 c_1^2
\end{eqnarray}
\end{scriptsize}
The determinant of the metric tensor turns out to be a rational
polynomial function in the $ \lbrace a_1, a_2 \rbrace $, which in
turn is given, in compact notations, as
\begin{scriptsize}
\begin{eqnarray}
g = -\frac{4 V^4 c_2^2 c_1^2}{R_0^2} \frac{n^C_g}{r^C_g}
\end{eqnarray}
\end{scriptsize}
Herewith, the numerator of the determinant of the metric tensor is
given by the following trigonometric polynomial:
\begin{scriptsize}
\begin{eqnarray}
n^C_g:&=& -25 c_2^4 c_1^6-18 c_2^3 c_1^5 s_1 s_2  -c_2^6-c_1^6+13
c_1^5 c_2^5 s_1 s_2 \nonumber \\&& -c_1^5 s_1+s_2 c_2^5+5 s_1
c_2^5 s_2 c_1 -18 c_1^3 c_2^5 s_1 s_2+5 c_2 c_1^5 s_1 s_2
\nonumber \\&& +12 c_1^2 c_2^6-25 c_1^4 c_2^6+12 c_1^6 c_2^2 +5
c_1^4 c_2 s_2-5 c_1 c_2^4 s_1  \nonumber \\&& -10 c_1^3 c_2^2
s_1-11 c_1^4 c_2^2 -11 c_2^4 c_1^2+10 c_1^2 c_2^3 s_2 -60 s_1
c_2^6 c_1^5 \nonumber \\&& +60 s_2 c_2^5 c_1^6 -20 c_1^6 c_2^3 s_2
+20 c_1^3 c_2^6 s_1 -2 c_1^5 c_2^2 s_1-47 s_2 c_2^5 c_1^4
\nonumber \\&& +2 s_2 c_2^5 c_1^2+47 c_1^5 c_2^4 s_1 +14 c_1^3
c_2^3 s_1 s_2+14 c_1^6 c_2^6  +36 c_1^4 c_2^4 \nonumber
\\&& -10 c_1^4 c_2^3 s_2 +10 c_1^3 c_2^4 s_1
\end{eqnarray}
\end{scriptsize}
As before, it can be seen that the denominator of the metric is
given by the following trigonometric polynomial:
\begin{scriptsize}
\begin{eqnarray}
r^C_g:&=& -210 c_2^4 c_1^6+252 c_1^5 c_2^5 s_1 s_2 +460 c_1^4
c_2^8-1190 c_1^6 c_2^8+460 c_1^8 c_2^4 \nonumber \\&& -1190 c_1^8
c_2^6-45 c_1^2 c_2^8-45 c_1^8 c_2^2 +45 c_1^2 c_2^{10}-250 c_1^4
c_2^{10}+490 c_1^6 c_2^{10} \nonumber \\&& -405 c_1^8 c_2^{10}-405
c_1^{10} c_2^8+490 c_1^{10} c_2^6 +45 c_1^{10} c_2^2-250 c_1^{10}
c_2^4 \nonumber \\&& -584 c_1^5 c_2^7 s_1 s_2 +120 c_1^7 c_2^3 s_1
s_2 +120 c_1^3 c_2^7 s_1 s_2 -584 c_1^7 c_2^5 s_1 s_2 \nonumber
\\&& +10 s_1 s_2 c_1^9 c_2 +10 s_1 c_2^9 s_2 c_1 -344 c_1^7 c_2^9 s_1 s_2
-120 c_1^9 c_2^3 s_1 s_2 \nonumber \\&& +332 c_1^9 c_2^5 s_1 s_2
-344 c_1^9 c_2^7 s_1 s_2 -120 s_1 c_2^9 s_2 c_1^3 +332 s_1 c_2^9
s_2 c_1^5 \nonumber \\&& +1180 c_1^8 c_2^8 -210 c_1^4
c_2^6-c_1^{10} +808 c_1^7 c_2^7 s_1 s_2 +910 c_1^6 c_2^6 \nonumber
\\&& -c_2^{10}+121 c_1^{10} c_2^{10} +122 c_1^9 c_2^9 s_1 s_2
\end{eqnarray}
\end{scriptsize}
We see that the determinant of the metric tensor remains non-zero
in the space of power factors and thus defines a non-degenerate
intrinsic geometry on the surface of fluctuation of phases.
\begin{figure}
\hspace*{0.5cm}
\includegraphics[width=8.0cm,angle=-90]{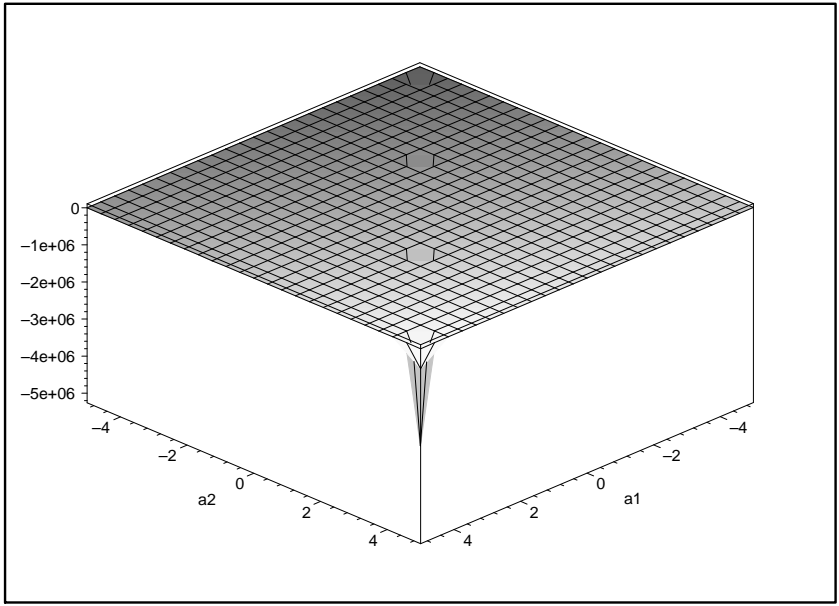}
\caption{The determinant of the metric tensor plotted as the
function of the power factors $a_1, a_2$, describing the complex
power fluctuations in electrical networks.}  \label{det3dcomplex}
\vspace*{0.5cm}
\end{figure}

\begin{figure}
\hspace*{0.5cm}
\includegraphics[width=8.0cm,angle=-90]{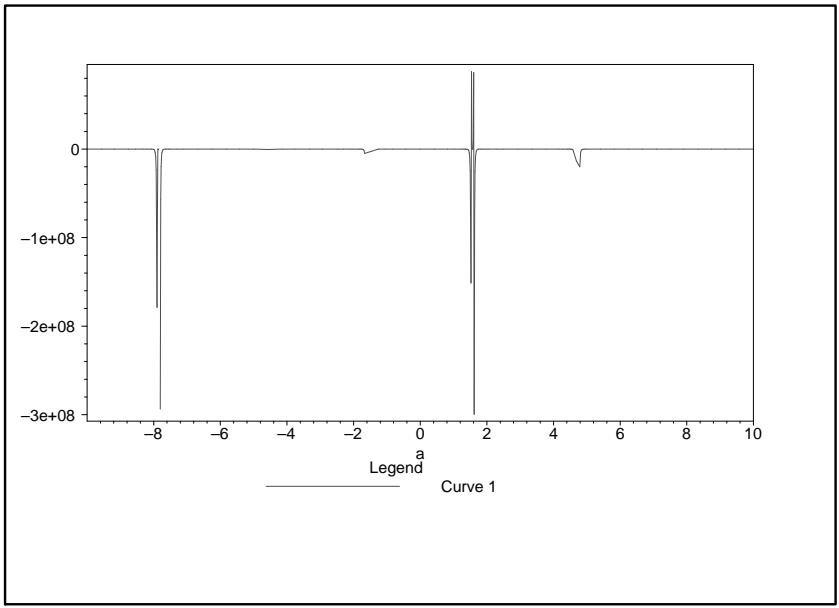}
\caption{The determinant of the metric tensor plotted as the
function of the equal power factor $a:= a_1= a_2$, describing the
complex power fluctuations in electrical networks.}
\label{det2dcomplex} \vspace*{0.5cm}
\end{figure}

Finally, we may easily obtain the underlying scalar curvature
which also has a rational polynomial form. It turns out that the
scalar curvature can be reduced to the following form:
\begin{scriptsize}
\begin{eqnarray}
R = \frac{R_0}{4c_1^2c_2^2V^2}
\frac{n^{(1)C}_R+n^{(2)C}_R+n^{(3)C}_R+n^{(4)C}_R}{r^{(1)C}_R+r^{(2)C}_R+r^{(3)C}_R}
\end{eqnarray}
\end{scriptsize}
Our computation shows that the numerator of the scalar curvature
takes the following trigonometric polynomial expressions:
\begin{scriptsize}
\begin{eqnarray}
n^{(1)C}_R&:=& -495 c_1^4 c_2^8-234 c_1^6 c_2^8 -495 c_1^8 c_2^4
-234 c_1^8 c_2^6 -66 c_1^2 c_2^{10}+162 c_1^4 c_2^{10} \nonumber
\\&& +13566 c_1^6 c_2^{10}-91736 c_1^8 c_2^{10} -91736 c_1^{10} c_2^8 +13566
c_1^{10} c_2^6 -66 c_1^{10} c_2^2\nonumber
\\&& +162 c_1^{10} c_2^4 +792 c_1^5 c_2^7 s_1 s_2+792 c_1^7 c_2^5 s_1 s_2 -21064 c_1^7
c_2^9 s_1 s_2+220 c_1^9 c_2^3 s_1 s_2 \nonumber \\&& +492 c_1^9
c_2^5 s_1 s_2-21064 c_1^9 c_2^7 s_1 s_2 +220 s_1 c_2^9 s_2
c_1^3+492 s_1 c_2^9 s_2 c_1^5  \nonumber \\&& +25506 c_1^8
c_2^8-c_1^{12}-c_2^{12}+2 c_2^{14} +1152 c_1^7 c_2^7 s_1 s_2 +2
c_1^{14} -924 c_1^6 c_2^6 \nonumber \\&& +175360 c_1^{11} c_2^{11}
s_1 s_2 +233534 c_1^{10} c_2^{10} +85296 c_1^9 c_2^9 s_1 s_2 +186
s_1 c_1^{13} c_2^2 \nonumber \\&&-1960 s_1 c_1^{13} c_2^4 +5512
s_1 c_1^{13} c_2^6 +3214 s_1 c_1^{13} c_2^8 -34758 s_1 c_1^{13}
c_2^{10} \nonumber \\&& +49084 s_1 c_1^{13} c_2^{12} -2054 s_2
c_1^4 c_2^{11} +10386 s_2 c_1^6 c_2^{11} -9732 s_2 c_1^8 c_2^{11}
\nonumber \\&& -33034 s_2 c_1^{10} c_2^{11} +72538 s_2 c_1^{12}
c_2^{11} -38450 s_2 c_1^{14} c_2^{11} +14436 s_2 c_1^8 c_2^9
\nonumber \\
n^{(2)C}_R&:=& 90 s_2 c_1^2 c_2^{11} -5796 s_2 c_1^8 c_2^7 +594
s_2 c_1^8 c_2^5 -3878 s_2 c_1^{10} c_2^5 \nonumber \\&& +14424 s_2
c_1^{10} c_2^7 -6944 s_2 c_1^{10} c_2^9 +6996 s_2 c_1^{12} c_2^5
-12292 s_2 c_1^{12} c_2^7 \nonumber \\&& -20348 s_2 c_1^{12} c_2^9
-3600 s_2 c_1^{14} c_2^5 +2428 s_2 c_1^{14} c_2^7 +18194 s_2
c_1^{14} c_2^9 \nonumber \\&& -14436 s_1 c_1^9 c_2^8 +6944 s_1
c_1^9 c_2^{10} +20348 s_1 c_1^9 c_2^{12} -18194 s_1 c_1^9 c_2^{14}
\nonumber \\&& -10386 s_1 c_1^{11} c_2^6 +9732 s_1 c_1^{11} c_2^8
+33034 s_1 c_1^{11} c_2^{10} -72538 s_1 c_1^{11} c_2^{12}
\nonumber \\&& +38450 s_1 c_1^{11} c_2^{14} -5074 s_2 c_1^6 c_2^9
-902 s_2 c_1^{12} c_2^3 +694 s_2 c_1^{14} c_2^3 \nonumber \\&&
+5074 s_1 c_1^9 c_2^6 +440 s_2 c_1^4 c_2^9 -30 s_2 c_1^{14} c_2
+2054 s_1 c_1^{11} c_2^4 \nonumber \\&& +5796 s_1 c_1^7 c_2^8
-14424 s_1 c_1^7 c_2^{10} -2428 s_1 c_1^7 c_2^{14} -440 s_1 c_1^9
c_2^4 \nonumber \\&& -660 s_1 c_1^7 c_2^6 +12292 s_1 c_1^7
c_2^{12} +902 s_1 c_1^3 c_2^{12} -6996 s_1 c_1^5 c_2^{12}
\nonumber \\
n^{(3)C}_R&:=& +3878 s_1 c_1^5 c_2^{10} +3600 s_1 c_1^5 c_2^{14}
-694 s_1 c_1^3 c_2^{14} +242 s_2 c_1^{10} c_2^3 \nonumber
\\&& -594 s_1 c_1^5 c_2^8 +660 s_2 c_1^6 c_2^7 -242 s_1 c_1^3
c_2^{10} -20 s_1 c_1 c_2^{12} \nonumber \\&& +30 s_1 c_1 c_2^{14}
+20 s_2 c_1^{12} c_2 -90 s_1 c_1^{11} c_2^2 -49084 s_2 c_1^{12}
c_2^{13} \nonumber \\&& +34758 s_2 c_1^{10} c_2^{13} +1960 s_2
c_1^4 c_2^{13} -3214 s_2 c_1^8 c_2^{13} -5512 s_2 c_1^6 c_2^{13}
\nonumber \\&& -186 s_2 c_1^2 c_2^{13} -32 s_1 s_2 c_1^3 c_2^{11}
-288 s_1 s_2 c_1^3 c_2^{13} -5720 s_1 s_2 c_1^5 c_2^{11} \nonumber
\\&& +5288 s_1 s_2 c_1^5 c_2^{13} +44672 s_1 s_2 c_1^7 c_2^{11}
-28816 s_1 s_2 c_1^7 c_2^{13}  \nonumber \\&& -132516 s_1 s_2
c_1^9 c_2^{11} +72372 s_1 s_2 c_1^9 c_2^{13} -132516 s_1 s_2
c_1^{11} c_2^9  \nonumber \\&&  -84592 s_1 s_2 c_1^{11} c_2^13
+72372 s_1 s_2 c_1^{13} c_2^9 -84592 s_1 s_2 c_1^{13} c_2^{11}
\nonumber \\&& -28816 s_1 s_2 c_1^{13} c_2^7 -32 s_1 s_2 c_1^{11}
c_2^3 -288 s_1 s_2 c_1^{13} c_2^3  \nonumber \\
n^{(4)C}_R&=& -5720 s_1 s_2 c_1^{11} c_2^5 +44672 s_1 s_2 c_1^{11}
c_2^7 +5288 s_1 s_2 c_1^{13} c_2^5  \nonumber \\&& +12 s_1 s_2
c_1^{11} c_2 -12 s_1 s_2 c_1^{13} c_2 -12 s_1 s_2 c_1 c_2^{13} +12
s_1 s_2 c_1 c_2^{11} \nonumber \\&& +14504 c_1^6 c_2^{14} -1552
c_1^4 c_2^{14}-26392 c_2^{12} c_1^6 +122669 c_2^{12} c_1^8
\nonumber \\&& -264682 c_2^{12} c_1^{10} +269177 c_2^{12} c_1^{12}
-102868 c_2^{12} c_1^{14} -26392 c_1^{12} c_2^6  \nonumber \\&&
+122669 c_1^{12} c_2^8 -264682 c_1^{12} c_2^{10}-1552 c_1^{14}
c_2^4 +14504 c_1^{14} c_2^6  \nonumber \\&& -56702 c_1^{14} c_2^8
+110054 c_1^{14} c_2^{10}-2 s_1 c_1^{13} +6 c_1^{14} c_2^2+6 c_1^2
c_2^{14} \nonumber \\&& +70 c_1^2 c_2^{12} +1771 c_1^{12} c_2^4
+70 c_1^{12} c_2^2+1771 c_1^4 c_2^{12} +2 s_2 c_2^{13} \nonumber
\\&& -102868 c_1^{12} c_2^{14} -56702 c_1^8 c_2^{14} +110054
c_1^{10} c_2^{14} +36560 s_1 s_2 c_1^{13} c_2^{13}  \nonumber
\\&& -21276 s_1 c_1^{13} c_2^{14} +21276 s_2 c_1^{14} c_2^{13}
+36556 c_1^{14} c_2^{14}
\end{eqnarray}
\end{scriptsize}
We see further that the denominator of the scalar curvature has
the following trigonometric polynomials:
\begin{scriptsize}
\begin{eqnarray}
r^{(1)C}_R&:=& -210 c_2^4 c_1^6+252 c_1^5 c_2^5 s_1 s_2 -68 c_1^4
c_2^8+3660 c_1^6 c_2^8 \nonumber \\&& -68 c_1^8 c_2^4 +3660 c_1^8
c_2^6 -45 c_1^2 c_2^8-45 c_1^8 c_2^2 \nonumber \\&& -31 c_1^2
c_2^{10}+991 c_1^4 c_2^{10} -5074 c_1^6 c_2^{10} +5257 c_1^8
c_2^{10} \nonumber \\&& +5257 c_1^{10} c_2^8-5074 c_1^{10} c_2^6
-31 c_1^{10} c_2^2+991 c_1^{10} c_2^4 \nonumber \\&& +172 c_1^5
c_2^7 s_1 s_2 +120 c_1^7 c_2^3 s_1 s_2 +120 c_1^3 c_2^7 s_1 s_2
+172 c_1^7 c_2^5 s_1 s_2 \nonumber \\&& +10 s_1 s_2 c_1^9 c_2 +10
s_1 c_2^9 s_2 c_1 +5916 c_1^7 c_2^9 s_1 s_2 +138 c_1^9 c_2^3 s_1
s_2 \nonumber \\&& -2132 c_1^9 c_2^5 s_1 s_2 +5916 c_1^9 c_2^7 s_1
s_2 +138 s_1 c_2^9 s_2 c_1^3 -2132 s_1 c_2^9 s_2 c_1^5 \nonumber
\\&& -10970 c_1^8 c_2^8 -210 c_1^4 c_2^6 -672 s_1 c_2^6 c_1^5
+672 s_2 c_2^5 c_1^6 -c_1^{10} \nonumber \\&& -4112 c_1^7 c_2^7
s_1 s_2+70 c_1^6 c_2^6 -c_2^{10}+6836 c_1^{11} c_2^{11} s_1
s_2+8265 c_1^{10} c_2^{10} \nonumber \\
r^{(2)C}_R&:=& -310 c_1^9 c_2^9 s_1 s_2-292 s_2 c_1^4 c_2^{11}
+2840 s_2 c_1^6 c_2^{11}-7846 s_2 c_1^8 c_2^{11} \nonumber \\&&
+8556 s_2 c_1^{10} c_2^{11}-3240 s_2 c_1^{12} c_2^{11} +8876 s_2
c_1^8 c_2^9-20 s_2 c_1^2 c_2^{11} \nonumber \\&& -220 s_2 c_1^8
c_2^7-1596 s_2 c_1^8 c_2^5 -120 s_2 c_1^{10} c_2^5+6560 s_2
c_1^{10} c_2^7 \nonumber
\\&& -14604 s_2 c_1^{10} c_2^9+1116 s_2 c_1^{12} c_2^5 -4784 s_2
c_1^{12} c_2^7+6942 s_2 c_1^{12} c_2^9  \nonumber \\&& -8876 s_1
c_1^9 c_2^8+14604 s_1 c_1^9 c_2^{10} -6942 s_1 c_1^9 c_2^{12}-2840
s_1 c_1^{11} c_2^6 \nonumber \\&& +7846 s_1 c_1^{11} c_2^8-8556
s_1 c_1^{11} c_2^{10} +3240 s_1 c_1^{11} c_2^{12}-688 s_2 c_1^6
c_2^9 \nonumber \\&& -24 s_2 c_1^{12} c_2^3+688 s_1 c_1^9 c_2^6
-618 s_2 c_1^4 c_2^9+292 s_1 c_1^{11} c_2^4 \nonumber \\&& +220
s_1 c_1^7 c_2^8-6560 s_1 c_1^7 c_2^{10} +618 s_1 c_1^9 c_2^4+2080
s_1 c_1^7 c_2^6 \nonumber \\&& +4784 s_1 c_1^7 c_2^{12}+24 s_1
c_1^3 c_2^{12} -1116 s_1 c_1^5 c_2^{12}+120 s_1 c_1^5 c_2^{10}
\nonumber \\
r^{(3)C}_R&:=&-284 s_2 c_1^{10} c_2^3+1596 s_1 c_1^5 c_2^8 -2080
s_2 c_1^6 c_2^7+284 s_1 c_1^3 c_2^{10} \nonumber \\&& +10 s_1 c_1
c_2^{12}-10 s_2 c_1^{12} c_2  +20 s_1 c_1^{11} c_2^2-258 c_1^3
c_2^8 s_1  \nonumber \\&& -20 c_1 c_2^{10} s_1+258 c_1^8 c_2^3 s_2
-492 c_1^7 c_2^4 s_1+492 c_1^4 c_2^7 s_2 \nonumber \\&& -196 s_1
s_2 c_1^3 c_2^{11}+748 s_1 s_2 c_1^5 c_2^{11} +768 s_1 s_2 c_1^7
c_2^{11}-6886 s_1 s_2 c_1^9 c_2^{11} \nonumber \\&& -6886 s_1 s_2
c_1^{11} c_2^9-196 s_1 s_2 c_1^{11} c_2^3  +748 s_1 s_2 c_1^{11}
c_2^5+768 s_1 s_2 c_1^{11} c_2^7 \nonumber \\&& +10 s_1 s_2
c_1^{11} c_2+10 s_1 s_2 c_1 c_2^{11} +794 c_2^{12} c_1^6+3366
c_2^{12} c_1^8 \nonumber \\&& -10303 c_2^{12} c_1^{10} +6835
c_2^{12} c_1^{12} +794 c_1^{12} c_2^6+3366 c_1^{12} c_2^8
\nonumber \\&& -10303 c_1^{12} c_2^{10}+53 c_1^2 c_2^{12} -489
c_1^{12} c_2^4 +53 c_1^{12} c_2^2 -489 c_1^4 c_2^{12}  \nonumber
\\&& +2 c_2^{11} s_2 -2 c_1^{11} s_1+92 c_1^2 c_2^9 s_2
+20 c_1^{10} c_2 s_2-92 c_1^9 c_2^2 s_1
\end{eqnarray}
\end{scriptsize}

\begin{figure}
\hspace*{0.5cm}
\includegraphics[width=8.0cm,angle=-90]{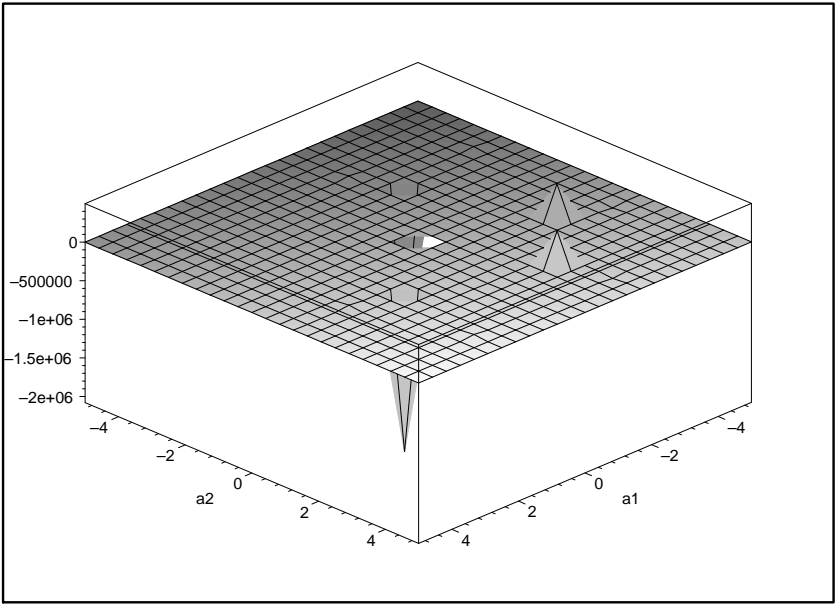}
\caption{The curvature scalar plotted as a function of the power
factors $a_1, a_2$, describing the complex power fluctuations in
electrical networks.}  \label{cur3dcomplex} \vspace*{0.5cm}
\end{figure}

\begin{figure}
\hspace*{0.5cm}
\includegraphics[width=8.0cm,angle=-90]{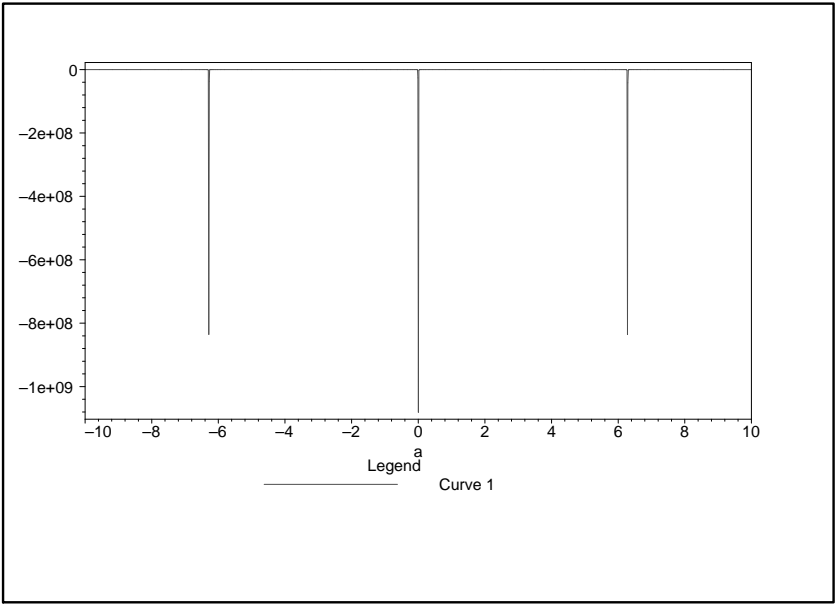}
\caption{The curvature scalar plotted as a function of the equal
power factor $a:= a_1= a_2$, describing the complex fluctuations
in electrical networks.}  \label{cur2dcomplex} \vspace*{0.5cm}
\end{figure}

For the choices $V=1$ and $R_0=1$, the Fig.(\ref{det3dcomplex})
shows that the determinant of the metric tensor has a commutative
effect of the real and complex power flow fluctuations. This plot
explicates the nature of the stability of a joint power flow in a
realistic electrical network. The corresponding plot for the
scalar curvature is depicted in the Fig.(\ref{cur3dcomplex}). This
plot shows the global nature of a combined real and imaginary
power flows in the electrical network. This analysis remains under
the effect of Gaussian fluctuations of the realistic network of
the resistance, reactance and impedance.

For the equal network parameters $a_1=a$ and $a_2=a$, the surface
plots of the determinant of the metric tensor and scalar curvature
are respectively shown in the Figs.(\ref{det2dcomplex}) and
(\ref{cur2dcomplex}). We observe herewith that the stability of
the electrical network power flow exists in certain distorted
bands. These distortions are strong enough, so that they are
capable to modulate the global properties of the fluctuation of
the power flow. Specifically, we notice, for equal values of the
network power factors, that the global instabilities exit for
three specific values.

\section{Conclusion and Remarks}
We have studied a power system planning, where the voltage
stability is the main concern. In this setup, advancements in the
electricity market, thus introduced, involve radical changes in
the structure of the power systems. In the electrical environment
of the power industry, we have considered an intrinsic geometric
model to make a power system non-linearly efficient. Our model
gives promising optimization criteria to select the optimal
network parameters. The robustness of the model has been
illustrated by variation(s) of the impedance angles, viz., phases.
For a set of impedance, resistance and capacitance, our
construction describes a definite stability character, with
respect to the power fluctuations of the network(s). As a function
of the trigonometric polynomials, we have demonstrated that the
canonical fluctuations can precisely be depicted without any
approximation.

In the present paper, we have analyzed the statistical
fluctuations in the real and imaginary power flows and thus
characterized the network configurations. Furthermore, for the
joint effect of the real and imaginary power equations, we have
shown that the intrinsic geometric notion offers a clear picture
of fluctuating network parameters. Such a configuration, as the
limit of an ensemble of the power factor fluctuations, reduces to
the specific electrical network. The present analysis does not
stop here, but it invades the nature of the underlying imaginary
power flow and explicates the global stability for the identical
power factors. Our study thus offers an appropriate network design
towards the stability of the existing linear optimization
techniques.

Finally, our proposition offers suitable tests towards the
planning of the network parameters of finite component power
systems. In our analysis, the intrinsic geometric model takes an
account of the non-linear effects, arising in the stochastic power
systems. A novel approach is herewith made possible in the history
of network designs, and thus the present investigation provides
appropriate network designs. The criteria deduced from our method
can be used for determining optimization constraints both in the
economic analysis of power systems, as well as for setting up the
operating point(s) of the power system, viz., network planning and
compensation techniques. This task is left for the future
investigation.

\section*{Acknowledgment}
N.G. thanks Prof. R. Shekhar and Prof. P. K. Kalra for their
continued guidance and support towards the completion of this
paper. B. N. T. thanks Prof. V. Ravishankar for support and
encouragement. The work of BNT is supported by postdoctoral
research fellowship of the \textit{``INFN-Laboratori Nazionali di
Frascati, Roma, Italy''}.

\end{document}